\documentclass[prb,superscriptaddress,twocolumn,preprintnumbers,a4,showpacs,floatfix]{revtex4}  
\usepackage{amsmath, amssymb, psfrag, tabls, dcolumn, bm, color}
\usepackage[sort&compress]{natbib}
\usepackage{graphicx}

\newcommand{\be}{\begin{equation}} 
\newcommand{\ee}{\end{equation}}
\newcommand{\bma}{\begin{displaymath}}
\newcommand{\ema}{\end{displaymath}}

\begin{document} 

\title{Effects of Bending on Raman-active Vibration Modes of Carbon Nanotubes}

\author{Sami Malola}
\address{Department of Physics, NanoScience Center, 40014 University of Jyv\"askyl\"a, Finland}

\author{Hannu H\"akkinen}
\address{Department of Physics, NanoScience Center, 40014 University of Jyv\"askyl\"a, Finland}
\address{Department of Chemistry, NanoScience Center, 40014 University of Jyv\"askyl\"a, Finland}

\author{Pekka Koskinen\footnote{Author to whom correspondence should be addressed.}}
\email{pekka.koskinen@phys.jyu.fi}
\address{Department of Physics, NanoScience Center, 40014 University of Jyv\"askyl\"a, Finland}


\date{\today}

\begin{abstract} 
We investigate vibration modes and their Raman activity of single-walled carbon nanotubes
that are bent within their intrinsic elastic limits. By implementing novel boundary 
conditions for density-functional based tight-binding, and using non-resonant bond polarization 
theory, we discover that Raman activity can be induced by bending. Depending on the degree of 
bending, high-energy Raman peaks change their positions and intensities significantly. These 
effects can be explained by migration of nodes and antinodes along tube circumference.
We discuss the challenge of associating the predicted spectral changes with experimental observations.
\end{abstract}

\maketitle

Technological applications of carbon nanotubes (CNTs) are based on their exceptional 
mechanical and electronic properties. Due to advances in fabrication and manipulation, 
CNTs have become one of the most prominent building blocks for nanoscale materials design.
Their electronic and mechanical properties can be used in numerous applications
for nanoelectronics, hydrogen and energy storage material, sensors, or high-strength composites.\cite{Science_297_787}

Among mechanical properties\cite{Qian_AMR_02}, vibrations are relevant in heat 
dissipation\cite{PRL_95_266803}, sensors\cite{Science_287_622} and nanotube 
identification\cite{PRB_71_075401}. Due to the large number of applications, vibrations have been 
investigated extensively. In particular, most experimental studies use Raman 
spectroscopy\cite{Rev1}, a method that is able to achieve even single nanotube 
resolution.\cite{Science_301_1354}

In practice, because CNTs are long, they bend. Bending is observed in isolated CNTs between 
electrodes\cite{PRL_99_045901}, or in ``paper''\cite{JEC_488_92}, ``forests''\cite{Science_282_1105},  
rings\cite{rings}, and composite systems\cite{Loos_ultramic_05} made out of CNTs.
 It appears that bending is ubiquitous in experiments---and challenging to 
study theoretically. Most previous theoretical Raman studies are for straight tubes, because modeling 
of bent systems has been computationally too expensive\cite{Rev1}. Modeling of bending with classical methods is straightforward, but has been used to study force moments and 
strains\cite{MandPofS_56_279}, buckling\cite{compB_39_202,Maiti_CPL_00} and other large-scale 
mechanical properties that result from rather high curvature.\cite{Qian_AMR_02}

In this work we investigate how vibrations and Raman spectra are affected when CNTs are 
bent \emph{slightly}, within their intrinsic elastic limits. To accomplish this, 
novel boundary conditions are introduced which allow quantum mechanical modeling of bending with 
computationally feasible system size. We show that vibrations undergo systematic changes that 
significantly alter the high-frequency Raman spectra. The spectral changes can be understood via simple physical principles.

We use density-functional based tight-binding (DFTB) method\cite{frauenheim_JPCM_02} to calculate 
forces, optimize systems\cite{bitzek_PRL_06}, and calculate vibrational eigenmodes. The method has 
been used successfully for vibrational analysis of carbon nanotubes, also related to Raman 
activity.\cite{jiang_PRB_05b,jiang_PRB_07} Raman spectra are calculated by non-resonant bond 
polarization method\cite{bp1,bp2}. This method has some restrictions and limits direct comparison 
between resonant experiments\cite{ebptvsres,marinopoulos_PRL_03}, but suffices for the scope of this paper. 
The details of our approach are given in Ref.~\onlinecite{vacancies_PRB_08}.

\begin{figure}[tb!]
\includegraphics[width=5.5cm]{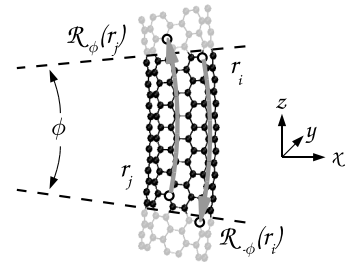}
\caption{Schematic of the boundary condition. Wave functions satisfy $\psi({\bf r}_{B})=\psi(\mathcal{R}_{\pm \phi}({\bf r}_{B}))$, where ${\bf r}_B$ is a point on either boundary (with one ${\bf k}$-point along tube axis); operation $\mathcal{R}_{\pm \phi}({\bf r}_j)$ rotates ${\bf r}_j$ an angle $\pm \phi$ around the wedge apex. We define the vector ${\bf r}_{ij}$ from atom $i$ to atom $j$ as equal to shortest of vectors $(\mathcal{R}_\phi({\bf r}_j)-{\bf r}_i)$, $(\mathcal{R}_{-\phi}({\bf r}_j)-{\bf r}_i)$ or $({\bf r}_j-{\bf r}_i)$, where ${\bf r}_i$ and ${\bf r}_j$ are atom positions within the simulation box. For some atom pairs ${\bf r}_{ij}$ crosses the boundary and renders Newton's third law invalid.  }
\label{fig1}
\end{figure}

To model bent nanotubes quantum mechanically, we introduce novel ``periodic wedge boundary conditions'' where the CNT appears as a slice of a torus, as shown in Fig.~\ref{fig1}. Some complications arise from fixed quantization axis and require mild approximations.\cite{wedge-comment} Henceforth all directions refer to the fixed Cartesian coordinates shown in Fig.~\ref{fig1}.

\begin{figure*}[tb!]
\includegraphics[width=15cm]{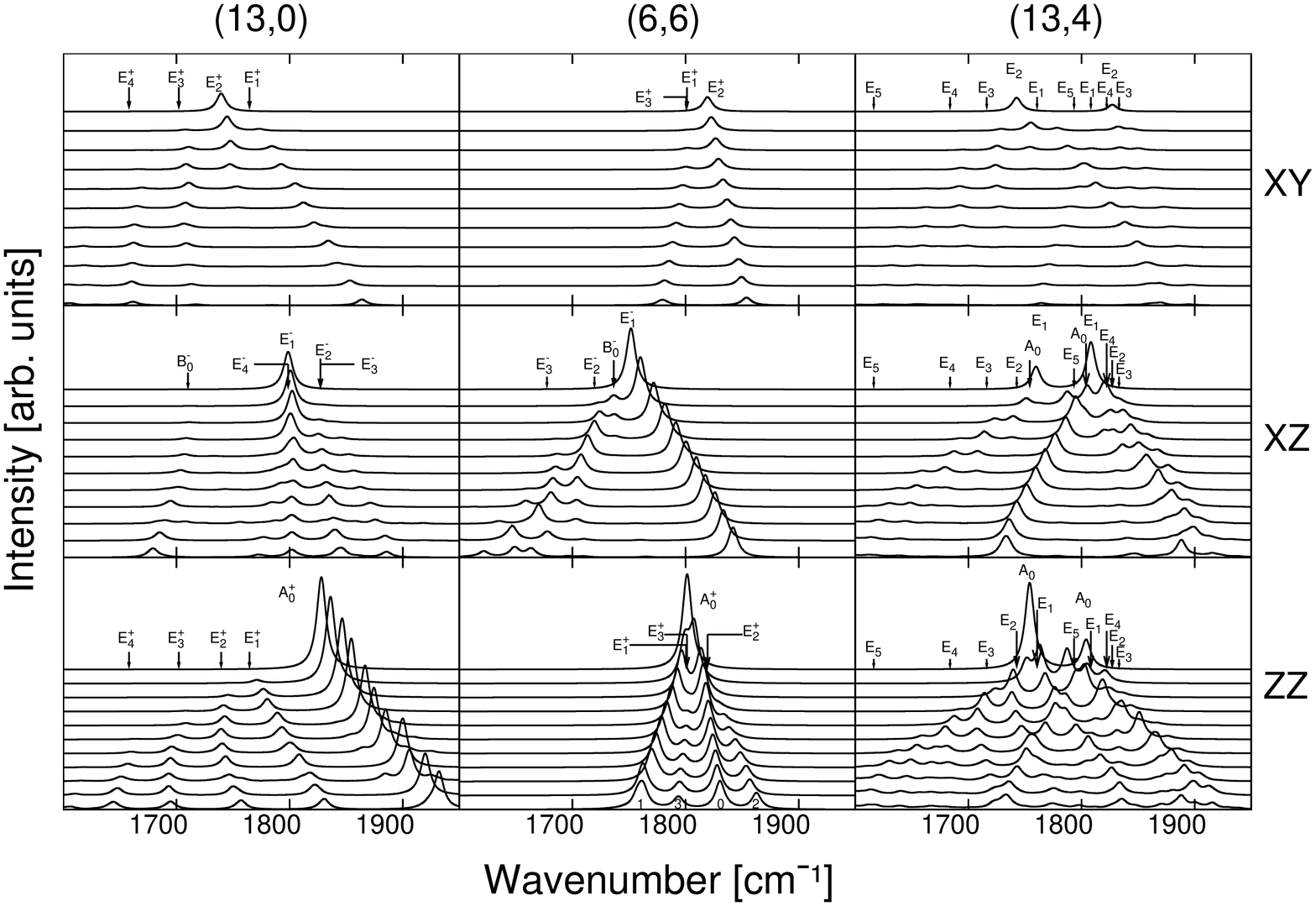}
\caption{Raman spectra of bent (13,0),(6,6) and (13,4) carbon nanotubes in different 
polarization pictures. Polarization picture is defined by directions of incident and scattered light as shown by symbols on the right. 
Bending increases linearly from zero (upmost lines) to $\Theta=4.2$~\% for (13,0), $\Theta=2.6$~\% for (6,6), and $\Theta=4.8$~\% for (13,4) tubes (lowest lines; corresponds to wedge angle $\phi=10^\circ$ for all tubes). The symbols with arrows refer to 
symmetry of modes for straight tubes in line group notation\cite{symaddC}. Peaks were Lorenzian broadened with full width 
at half maximum of $5$~cm$^{-1}$. Small symbols in the spectra of (6,6) tube with zz-polarization refer to the subscripts of original mode symmetries.}
\label{fig2}
\end{figure*}

We stress that in our calculations the bending of CNTs is \emph{only} due to boundary 
conditions. All atoms are free to move in the unit cell; no  constraints are applied
(those would cause severe artifacts for the vibration modes). Note that the length of the tube is automatically optimized since the atoms can freely move in the radial direction of the unit cell.

%
%
The single-walled CNTs we investigate are: semiconducting (13,0) tube with $10.2$~\AA~diameter and $21.3$~\AA~length, metallic (6,6) tube with $8.1$~\AA~diameter and $27.1$~\AA~length, and metallic (13,4) tube with $12.1$~\AA~diameter and $21.9$~\AA~length. To have a common measure for the degree of bending for tubes with different chiralities and diameters, we define a 
dimensionless variable
\[
\Theta=D/2R,
\]
where $D$ is tube's diameter and $R$ the radius of curvature measured from tube axis. Hence $\Theta=1.0$ (100 \%) corresponds 
to maximum bending (torus with a vanishing hole); buckling of a nanotube takes place above $\Theta \sim 10\;\%$\cite{compB_39_202,Maiti_CPL_00} and the range for bendings in experiments is estimated $\Theta=0.05\;\% \ldots 
5\;\%$\cite{Loos_ultramic_05,PRL_99_045901}, which is the range under our focus. 

Our main results are embedded in Fig.~\ref{fig2}, showing Raman spectra of high-energy modes by 
systematically varying $\Theta$. Low-energy modes are not shown because they are 
insensitive to bending, with respect to both energy and  intensity. For example, the radial 
breathing mode is left nearly intact because bending mostly affects bonds parallel to tube axis. On the 
contrary, within G-band -related high-energy modes bending induces systematic changes: (\emph{i}) emergence of new 
peaks, (\emph{ii}) intensity reductions of ``original'' (straight-tube) peaks, (\emph{iii}) 
significant energy shifts and (\emph{iv}) splitting of peaks into smaller ones. In (13,0) 
tube $E_2^+$ mode even deactivates for moderate bending. The rich spectra of chiral (13,4) tube is 
further enriched while bent, but still shows similar systematics to achiral tubes.

\begin{figure}[tb!]
\includegraphics[width=9cm,height=8cm]{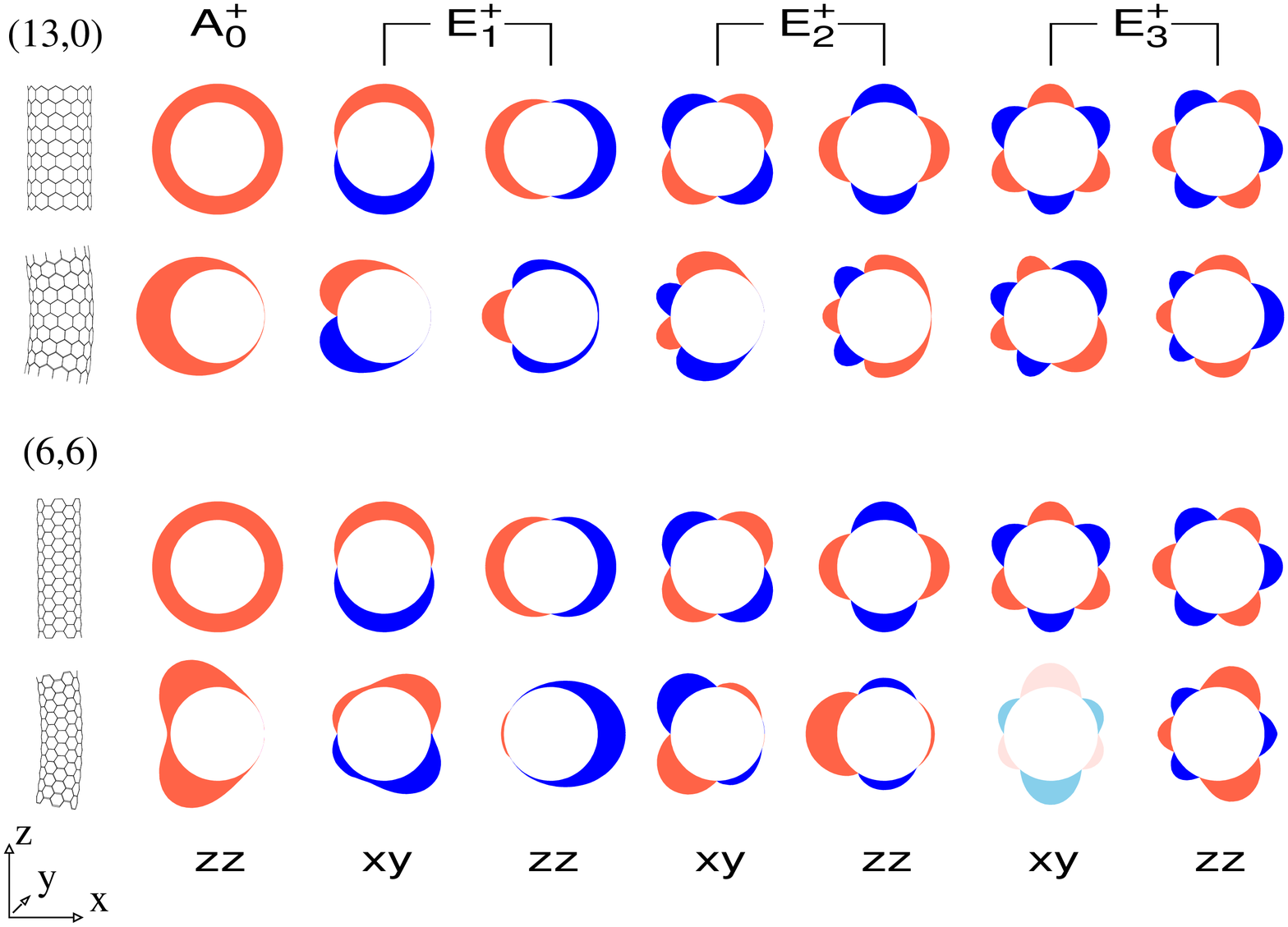}
\caption{(color online) Qualitative view of the nodes and antinodes of selected vibration modes 
for (13,0) and (6,6) tubes. Thick blue line represents an antinode, red an antinode with opposite phase and vanishing line a node; the actual direction of the vibration amplitude is either circumferential or along tube axis. Modes are for straight or bent tubes as indicated on the left. The symbols below show the polarization pictures where the bent modes become most visible. We were unable to identify one $E_3^+$ mode for (6,6) tube (faint symbol).}
\label{fig3}
\end{figure}

We begin analyzing these results by looking at what happens in the spatial structure of the vibration modes as tubes are bent. To visualize this, we show a qualitative view of the nodes and antinodes for selected modes in Fig.~\ref{fig3}.
Two-dimensional $E$ modes within the same polarization picture are similar and uniform in the
direction of the tube, differing only in the number of nodes along tube circumference. In xy and zz 
polarizations the amplitudes are in (13,0) tube along tube axis and in (6,6) tube along tube 
circumference (for xz vice versa). As the tube is bent, the nodal structure starts migrating 
towards the outer or inner side of the torus. Amplitudes are also affected, but the modes remain 
uniform along tube axis and the transition is smooth. In other words, modes in slightly bent tubes can be always
identified with the symmetric modes in straight tubes.

\begin{figure}[tb!]
\includegraphics[width=8cm]{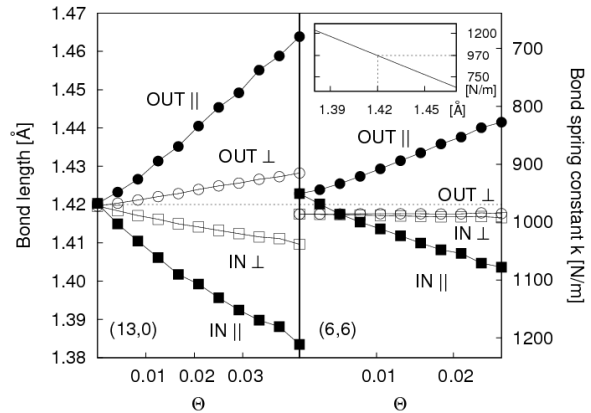}
\caption{Variation of bond lengths for (13,0) (left) and (6,6) (right) tubes as a function of 
bending parameter $\Theta$. Bonds are divided into mostly parallel ($||$) and mostly perpendicular 
($\bot$) bonds with respect to tube axis (in and out refer to the sides of torus). Inset shows the 
effective nearest neighbor spring constant, calculated by scaling single graphene layer; the inset 
was used to map the bond length scale (left-hand axis) into spring constant scale (right-hand 
axis). The equilibrium spring constant $970$~N/m and thermal expansion coefficient $5.13\cdot 
10^{-6}$~K$^{-1}$, used to make the inset, are in overall agreement with reference 
results.\cite{Riter_JCP_70,Schelling_PRB_03}}
\label{fig4}
\end{figure}

A qualitative change for bent tubes, as shown in Fig.~\ref{fig4} as a function 
of $\Theta$, is the change in bond lengths and bond stiffnessess. The bonds in inner parts shorten, bonds in outer parts lengthen, with up to $5$~\% 
variation for $\Theta=4$~\%. Because carbon bonds in honeycomb structure are stiff, the result 
shows a surprisingly strong ($\sim 50$~\%) variation in the effective (nearest neighbor) 
spring constants between the inner and outer parts of the tube. 

These observations help us to understand the energy shifts, remembering that $\omega \sim 
\sqrt{k}$. Let us take the $zz$-polarization spectrum of (6,6) tube in Fig.~\ref{fig2} as an example. The 
energies of original $A_0^+$ and $E_2^+$ symmetry modes increase; it is because the antinodes 
concentrate in the inner part of the torus where the spring constants are larger (see 
Fig.~\ref{fig3}). The energies of vibrations originating from $E_1^+$ and $E_3^+$ symmetry modes 
decrease; it is because the antinodes concentrate in the outer part of the torus where the spring 
constants are smaller (see Fig.~\ref{fig3}). Furthermore, looking at Fig.~\ref{fig3} one can 
realize that the energy of $E_1^+$ mode decreases more than $E_3^+$ because the antinodes of 
$E_1^+$ concentrate in outer part more strongly. This is part of a more general trend observable in 
Fig.~\ref{fig3}: modes with many circumferential nodes are less affected because antinodes are more 
equally distributed as tubes are bent. The same observations apply to other spectra. Note that in 
Fig.~\ref{fig3} bending modifies two-dimensional modes pairwise the same way (antinodes migrate in 
same direction), and bending does not lift the degeneracy of the modes because energy is for both 
modes either increased or decreased. Hence the appearance of peak splittings in Fig.~\ref{fig2} is 
due to originally different modes, not due to lifted degeneracy.

Why does bending cause migration of nodes and antinodes along the circumference? Consider a simple 
model: calculate the eigenmodes of a simple, one-dimensional ring of atoms connected by harmonic 
springs, where spring constants are modulated such that on one side of the ring they are larger, on 
the other side smaller. For this modified system the symmetric eigenmodes change so that nodes and 
antinodes concentrate either on the region of strong bonds, or on the region of weak bonds, 
depending on the mode in question. In nanotubes the modes and amplitudes are three-dimensional, but 
the basic mechanism remains the same.   

Finally, let us investigate the induced Raman activity of Fig.~\ref{fig2}. Consider for example the $E_1^+$ mode of (6,6) tube in Fig.~\ref{fig3} that is originally Raman inactive with xy and zz 
polarizations. Bending breaks the symmetry in $x$-direction and causes $E_1^+$ mode to resemble 
$A_0^+$ and $E_2^+$ modes---modes that are Raman active for straight tubes. Due to this resemblance 
$E_1^+$ and $A_0^+$ get nearly equal Raman intensities\cite{vacancies_PRB_08}. More generally, from 
$y$-antisymmetric modes, that can be excited by $y$-polarized light but that have broken 
$x$-symmetry, some will ``leak'' the polarization into $x$-direction, making the mode active with 
$xy$ polarization (e.g. left-hand $E_1^+$'s for (6,6) tube in Fig.~\ref{fig3}). Further, from 
$y$-symmetric modes, that have broken $x$-symmetry, some become active in $zz$ polarization (e.g. 
right-hand $E_1^+$'s for (6,6) tube in Fig.~\ref{fig3}). For (13,4) tube similar principles apply, but corresponding analysis is somewhat more complicated due to the chirality of the tube.

The arguments above are biased towards the non-resonant bond polarizability model, but we stress 
that most arguments are related to vibrational eigenmodes (peak positions, node migration), and are 
independent of the method to calculate the Raman intensities. Therefore most observations should be 
consistent with more complete theories. 

In fact, even the most complete theory would have problems with experimental interpretation. 
Spectra are calculated only for a piece of potentially complex curved CNT system. We confirmed 
that the tiny $x$-component, that tube axis has near boundaries, is not the origin for Raman 
activity ``leakage'' between polarization pictures. But if nanotube slice is a part of more 
complete torus, situation becomes more complicated and direct comparison less sensible, because 
tube axis mixes with other directions. Because the situation depends crucially on the experimental 
setups, we cannot make here general interpretations. 

There have been few experiments aiming for direct observation of Raman spectra for bent 
nanotubes\cite{rings,ko_APL_04}. In Ref.~\onlinecite{ko_APL_04} the $G$ band peak was observed to broaden 
and shift lower in energy, which was attributed to the increased bond lengths in bent tubes. 
Because for ideally bent tubes some modes should also decrease in energy, it is likely that in this 
experiment the tubes were not only bent but also stretched; shift is due to stretching and 
broadening due to bending. Our highest bending limit, $\Theta=5$~\%, was derived from 
microscopic images\cite{Loos_ultramic_05}, but high density of such bendings must involve defects. It is 
because a bent tube must be anchored via mechanical or chemical bonds; a rough estimate for the 
bending energy of a tube with any chirality is $E_{\textrm{bend}}\approx 20\cdot 
\Theta^2$~(eV/atom), which for $\Theta=5$~\% bending requires significant $5$~eV of anchoring 
energy per $100$ atoms in the tube. Therefore defect-free extreme bending should occur only in singular parts of compound structures, and the Raman intensity from these parts is expected to remain comparably small. On the other hand, Raman measurements for single CNTs could show these effects visibly, especially under experimentally feasible controlled bending\cite{PRL_99_045901}.

To conclude, this work provides understanding into effects that
bending causes for vibrational spectra of CNTs. The effects are significant, but can be 
systematically explained with general principles, which should be valid for multi-walled nanotubes 
or even for non-carbon nanotubes.

This research is supported by the Academy of Finland through projects 121701 and 117997, the FINNANO consortium MEP (molecular electronics and nanoscale photonics), and by the Finnish Cultural Foundation (SM). We thank  M. Pettersson and M. Manninen for fruitful discussions.
The computations were done in the NanoScience Center (NSC), University of Jyv\"askyl\"a. 


\begin{thebibliography}{29}
\bibitem[{wed()}]{wedge-comment}
\bibinfo{note}{Fixed quantization axis complicates matrix element calculation
  for some orbital pairs that cross a boundary. This requires averaged matrix
  elements and forces between boundaries. By using long tubes, small $\phi$'s,
  and avoiding atoms too near wedge apex, these approximations are valid.}
\end{thebibliography}

\end{document}